\documentstyle[12pt]{article}
\begin{document}
\def\bea{\begin{eqnarray}}
\def\eea{\end{eqnarray}}
\title{\bf {Casimir Forces for Robin Scalar Field on Cylindrical Shell in de Sitter Space }}

\author{M. R. Setare  \footnote{E-mail: rezakord@ipm.ir }
 \\
 {Institute for Theoretical
Physics and Mathematics}, \\
{P.O. Box 19395-5531, Tehran, Iran}
\\ {Department of Science, Physics Group, Kurdistan University,
Sanandeg, Iran}}

\maketitle
\begin{abstract}
The Casimir stress on a cylinderical shell in background of
conformally flat space-time for massless scalar field is
investigated. In the general case of Robin (mixed) boundary
condition formulae are derived for the vacuum expectation values
of the energy-momentum tensor and vacuum forces acting on
boundaries. The special case of the dS bulk is considered then
different cosmological constants are assumed for the space inside
and outside of the shell to have general results applicable to the
case of cylindrical domain wall formations in the early universe.

 \end{abstract}
\newpage

 \section{Introduction}
The Casimir effect is one of the most interesting manifestations
  of nontrivial properties of the vacuum state in quantum field
  theory [1,2]. Since its first prediction by
  Casimir in 1948 \cite{Casimir} this effect has been investigated for
  different fields having different boundary geometries[4-7]. The
  Casimir effect can be viewed as the polarization of
  vacuum by boundary conditions or geometry. Therefore, vacuum
  polarization induced by a gravitational field is also considered as
  Casimir effect. The types of boundary and conditions that have been most often
  studied are those associated to well known problems, e.g. plates, spheres, and
  vanishing conditions, perfectly counducting conditions, etc. The cylindrical problem with
  perfectly counducting conditions was first considered in \cite{Mil}, for recent study
  ref.\cite{{Gos},{Mil1}}.   \\
  In the context of hot big bang cosmology, the unified theories
  of the fundamental interactions predict that the universe passes
  through a sequence of phase transitions. These phase transitions
  can give rise to domain wall structures determined by the topology of
  the manifold $M$ of degenerate vacuua \cite{{zel},{kib},{viel}}.
  If $M$ is disconnected, i.e. if $\pi(M)$ is nontrivial, then one
  can pass from one ordered phase to the other only by going
  through a domain wall. If $M$ has two connected components, e.g.
  if there is only a discrete reflection symmetry with
  $\pi_{0}(M)=Z_{2}$, then there will be just two ordered phase
  separated by a domain wall.\\
  The time evolution of topological defects have played an
  important role in many branches of physics, e.g., vortices in
  superconductors \cite{hu} and in superfluid \cite{do}, defects
  in liquid crystals \cite{ch}, domain wall \cite{{ar1},{ar2}},
  cosmic string \cite{{kib},{viel}} and a flux tube in QCD
  \cite{ba}.\\
  Zeldovich et al \cite{zel} have been shown that the energy
  density of the domain walls is so large that they would dominate
  the universe completely, violating the observed approximation isotropy and
  homogeneity. In other words, the domain walls were assumed to
  somehow disappear again soon after their creation in the early
  universe, for instance, by collapse, evaporation, or simply by inflating away
  from our visible universe. Much later however Hill et al\cite{hil} introduced
  the so called light or soft domain walls. They considered a
  late-time phase transition and found that light domain walls
  could be produced, that were not necessarily in contradication
  with observed large-scale structure of the universe. In addition, whatever the cosmological
  effects are, we find it important to obtain a better
  understanding of the dynamics of domain walls.\\
  Casimir effect in curved space-time has not been studied extensively.
  Casimir effect in the presence of a general relativistic domain
  wall is considered in \cite{set} and a study of the relation
  between trace anomaly and the Casimir effect can be found in
  \cite{set1}. Casimir effect may  have interesting implications for the early
  universe. It has been shown, e.g., in\cite{Ant} that a closed Robertson-Walker
  space-time in which the only contribution to the  stress tensor comes from Casimir energy
   of a scalar field is excluded. In inflationary models, where the
   dynamics of bubbles may play a major role, this dynamical
   Casimir effect has not yet been taken into account. Let us
   mention that in \cite{set2} we have investigated the
   Casimir effect of a massless scalar field with Dirichlet
   boundary condition in spherical shell having different vacuua
   inside and outside which represents a bubble in early universe with
   false/true vacuum inside/outside. In this reference the sphere
   have zero thickness. In another paper \cite{set3} we have extended the
   analysis to the spherical shell with nonvanishing thickness.
    Parallel plates  immersed in different de Sitter
  spaces in- and out-side is calculated in \cite{set4}.  \\
  In the present paper we will investigate the vacuum expectation
values of the energy--momentum tensor of the conformally coupled
scalar field on background of the conformally flat space-time. We
will consider a cylindrical shell and boundary conditions of the
Robin type on the shell. The latter includes the Dirichlet and
Neumann boundary conditions as special cases. The Casimir
energy-momentum tensor for these geometries can be generated from
the corresponding flat spacetime results by using the standard
transformation formula. Then we consider cylindrical shell with
  constant comoving radius having different vacuums inside and outside, i.e. with
   false/true vacuum  inside/outside. Our model may be used to study the effect of
   the Casimir force on the dynamics of the cylindrical domain wall appearing in
  the simplest Goldston model. In this model potential of the scalar field
  has two equal minima corresponding to degenerate vacuua. Therefore, scalar field
  maps points at spatial infinity in physical space nontrivially into the
  vacuum manifold \cite{vil1}. Domain wall structure occur at the boundary between
  these regions of space. One may assume that the outer regions of cylinder are
  in $\Lambda_{out}$ vacuum corresponding to degenerate vacuua in domain wall
  configuration.\\
The Casimir effect for the general Robin boundary conditions on
background of the Minkowski spacetime was investigated in Ref.
\cite{RomSah} for flat boundaries, and in \cite{Saha01a,Saha01b}
for spherically and cylindrically symmetric boundaries in the case
of a general conformal coupling. Here we use the results of Ref.
\cite{Saha01b} to generate vacuum energy--momentum tensor for the
cylindrical shell in conformally flat backgrounds. The paper is
organized as follows. In the next section the vacuum expectation
values of the energy--momentum tensor and vacuum forces acting on
shell are evaluated for a general case of a conformally-flat
background. In section $3$ we study the bulk Casimir effect for a
conformal scalar when the bulk is a 4-dimensional de Sitter space.
 Finally, the results are re-mentioned and discussed in last
section

\section{Vacuum expectation values for the energy-momentum tensor }
In this paper we will consider a conformally coupled massless scalar field $%
\varphi (x)$ satisfying the equation
\begin{equation}
\left( \nabla _{\mu }\nabla ^{\mu }+\xi R\right) \varphi
(x)=0,\quad \xi =\frac{D-1}{4D}  \label{fieldeq}
\end{equation}
on background of a $D+1$--dimensional conformally flat spacetime
with the metric
\begin{equation}
g_{\mu \nu }=e^{-2\sigma (r)}\eta _{\mu \nu },\quad \mu ,\nu
=0,1,\ldots ,D. \label{metric}
\end{equation}
In Eq. (\ref{fieldeq}) $\nabla _{\mu }$ is the operator of the
covariant derivative, and $R$ is the Ricci scalar for the metric
$g_{\mu \nu }$. Note that for the metric tensor from Eq.
(\ref{metric}) one has
\begin{equation}
R=De^{2\sigma }\left[ 2\sigma ^{\prime \prime }-(D-1)\sigma ^{\prime 2}%
\right] ,  \label{Riccisc}
\end{equation}
where the prime corresponds to the differentiation with respect to
$r$. We will assume that the field satisfies the mixed boundary
condition
\begin{equation}
(A+Bn^{i}\nabla _{i})\varphi (x)=0  \label{robbc}
\end{equation}
on the cylindrical shell with radius $a$. Here $n^{i}$ is the
normal to the boundary surface, $\nabla _{i}$ - is the covariant
derivative operator, $A$ and $B$ are constants. The results in the
following will depend on the ratio of these coefficients only.
However, to keep the transition to the Dirichlet and Neumann cases
transparent we will use the form (\ref{robbc}).\\
It can be shown that for a conformally coupled scalar by using
field equation (\ref{fieldeq}) the expression for the
energy--momentum tensor can be presented in the form
\begin{equation}
T_{\mu \nu }=\nabla _{\mu }\varphi \nabla _{\nu }\varphi -\xi \left[ \frac{%
g_{\mu \nu }}{D-1}\nabla _{\rho }\nabla ^{\rho }+\nabla _{\mu
}\nabla _{\nu }+R_{\mu \nu }\right] \varphi ^{2},  \label{EMT1}
\end{equation}
where $R_{\mu \nu }$ is the Ricci tensor. The quantization of a
scalar filed on background of metric (2) is standard. Let
$\{\varphi _{\alpha }(x),\varphi _{\alpha }^{\ast }(x)\}$ be a
complete set of orthonormalized positive and negative frequency
solutions to the field equation (\ref {fieldeq}), obying boundary
condition (\ref{robbc}). By expanding the field operator over
these eigenfunctions, using the standard commutation rules and the
definition of the vacuum state for the vacuum expectation values
of the energy-momentum tensor one obtains
\begin{equation}
\langle 0|T_{\mu \nu }(x)|0\rangle =\sum_{\alpha }T_{\mu \nu }\{\varphi {%
_{\alpha },\varphi _{\alpha }^{\ast }\}},  \label{emtvev1}
\end{equation}
where $|0\rangle $ is the amplitude for the corresponding vacuum
state, and the bilinear form $T_{\mu \nu }\{{\varphi ,\psi \}}$ on
the right is determined by the classical energy-momentum tensor
(\ref{EMT1}). In the problem under consideration we have a
conformally trivial situation: conformally invariant field on
background of the conformally flat spacetime. Instead of
evaluating Eq. (\ref{emtvev1}) directly on background of the
curved metric, the vacuum expectation values can be obtained from
the corresponding flat spacetime results for a scalar field
$\bar{\varphi}$ by using the conformal properties of the problem
under consideration. Under the
conformal transformation $g_{\mu \nu }=\Omega ^{2}\eta _{\mu \nu }$ the $%
\bar{\varphi}$ field will change by the rule
\begin{equation}
\varphi (x)=\Omega ^{(1-D)/2}\bar{\varphi}(x),  \label{phicontr}
\end{equation}
where for metric (\ref{metric}) the conformal factor is given by
$\Omega =e^{-\sigma (r)}$. The boundary conditions for the field
$\bar{\varphi}(x)$ we will write as following
\begin{equation}
\left( \bar{A}+\bar{B}\partial _{r}\right) \bar{\varphi}%
=0,  \label{bounconflat}
\end{equation}
with constant Robin coefficients $\bar{A}$ and $\bar{B}$.
Comparing to the boundary conditions (\ref{robbc}) and taking into
account transformation rule (\ref{phicontr}) we obtain the
following relations between the corresponding Robin coefficients
\begin{equation}
\bar{A}=A+\frac{D-1}{2}\sigma ^{\prime }(a)e^{\sigma (a)}B,\quad
\bar{B}=Be^{\sigma (a)}. \label{coefrel}
\end{equation}
Note that as Dirichlet boundary conditions are conformally
invariant the Dirichlet scalar in the curved bulk corresponds to
the Dirichlet scalar in a flat spacetime. However, for the case of
Neumann scalar the flat spacetime counterpart is a Robin scalar
with $\bar{A}=(D-1)\sigma '(a)/2$ and $\bar{B}=1$. The Casimir
effect with boundary conditions (\ref{bounconflat}) on cylindrical
shell on background of the Minkowski spacetime is investigated in
Ref. \cite{RomSah}for a scalar field with a general conformal
coupling parameter. In the case of a conformally coupled scalar
the corresponding regularized VEV's for the energy-momentum tensor
have the form
\begin{equation}
\langle 0|T_{\mu \nu }|0\rangle ={\rm diag}(\varepsilon
,-p_{1},-p_{2},-p_{3},\ldots ,-p_{D}).  \label{emtdiag}
\end{equation}
Here $\varepsilon $ is the vacuum energy density, $p_{1}$, $p_{2}$, $%
p_{3}=p_{4}=\cdots =p_{D}$ are effective pressures in the radial,
azimuthal and longitudinal directions, respectively (vacuum
stresses). These quantities are determined by the relations
\begin{equation}
q_{SUB}=\frac{2^{1-D}\pi ^{-(D+1)/2}}{a^{D+1}\Gamma (D/2-1/2)}%
\sum_{n=-\infty }^{+\infty }\int_{0}^{\infty }dz\,
z ^{D+3}\frac{\bar{K}_{n}(z)}{\bar{I}_{n}(z)}%
F_{n}^{(q)}[I_{n}(zr/a)],  \label{qinsub}
\end{equation}
where $I_{n}(z)$ and $K_{n}(z)$ are the modified Bessel functions,
and
\begin{eqnarray}
F_{n}^{(\varepsilon )}[f(z)] &=&\frac{1}{D-1} f^{2}(z)+\left( 2\xi
-\frac{1}{2}\right) \left[ f^{\prime
2}(z)+\left( \frac{n^{2}}{z^{2}}+1\right) f^{2}(z)\right]  \label{Fnepsin} \\
F_{n}^{(p_{1})}[f(z)] &=&\frac{1}{2}\left[ \left( \frac{n^{2}}{z^{2}}%
+1\right) f^{2}(z)-f^{\prime 2}(z)\right] -\frac{2\xi
}{z}f(z)f^{\prime }(z)
\label{Fnp1in} \\
F_{n}^{(p_{2})}[f(z)] &=&-\left( 2\xi -\frac{1}{2}\right) \left[
f^{\prime
2}(z)+\left( \frac{n^{2}}{z^{2}}+1\right) f^{2}(z)\right] +\frac{2\xi }{z}%
f(z)f^{\prime }(z)-\frac{n^{2}}{z^{2}}f^{2}(z)  \label{Fnp2in} \\
F_{n}^{(p_{i})}[f(z)] &=&-F_{n}^{(\varepsilon )}[f(z)],\quad
i=3,\ldots ,D. \label{Fnpiin}
\end{eqnarray}
Here and below we use the notation
\begin{equation}
\bar{f}(z)\equiv Af(z)+(B/a)zf^{\prime }(z)  \label{fbar}
\end{equation}
for a given function $f(z)$. Similarly the vacuum expectation
values for the exterior of a single cylindrical shell can be
obtain, the result is as following \cite{RomSah}
\begin{equation}
q_{SUB}=\frac{2^{1-D}\pi ^{-(D+1)/2}}{a^{D+1}\Gamma
(D/2-1/2)}\sum_{n=-\infty }^{+\infty }\int_{0}^{\infty }dz\,z
^{D+3}
\frac{\bar{I}_{n}(z)}{\bar{K}_{n}(z)}%
F_{n}^{(q)}[K_{n}(zr/a)],  \label{qsubout}
\end{equation}
where we use notations (\ref{Fnepsin})-(\ref{Fnpiin}). As we see,
these quantities can be obtained from the ones for interior region
by the replacements $I\rightarrow K$, $K\rightarrow I$. Using the
expressions for the interior and exterior quantities we have
\begin{equation}
\begin{array}{c}
F=\frac{-2^{-D}\pi^{-(D+1)/2}}{a^{D+1}\Gamma(D/2-1/2)}\sum_{n=-\infty
}^{+\infty }\int_{0}^{\infty }dz\, z^{D+1} \times
\\
\times \left[ 2\beta
-4\xi +(z^{2}+n^{2}-\beta ^{2}+4\xi \beta )\frac{\left( \tilde{I}_{n}(z)%
\tilde{K}_{n}(z)\right) ^{\prime }}{z\tilde{I}_{n}^{\prime }(z)%
\tilde{K}_{n}^{\prime }(z)}\right]   \label{Fgum}
\end{array}
\end{equation}
for the total vacuum force acting per unit surface of the shell.
In these formulae we have introduced the notation
\begin{equation}
\tilde{f}(z)=z^{\beta }f(z),\qquad \beta =A/B  \label{nottilde}
\end{equation}
for a given function $f(z)$.
\\
The vacuum energy-momentum tensor on curved background
(\ref{metric}) is obtained by the standard transformation law
between conformally related problems (see, for instance,
\cite{davies}) and has the form
\begin{equation}
\langle T_{\nu }^{\mu }\left[ g_{\alpha \beta }\right] \rangle _{{\rm ren}%
}=\langle T_{\nu }^{\mu }\left[ g_{\alpha \beta }\right] \rangle _{{\rm ren}%
}^{(0)}+\langle T_{\nu }^{\mu }\left[ g_{\alpha \beta }\right] \rangle _{%
{\rm ren}}^{(b)}.  \label{emtcurved1}
\end{equation}
Here the first term on the right is the vacuum energy--momentum
tensor for the situation without boundaries (gravitational part),
and the second one is due to the presence of boundaries. As the
quantum field is conformally coupled and the background spacetime
is conformally flat the gravitational part of the energy--momentum
tensor is completely determined by the trace anomaly and is
related to the divergent part of the corresponding effective
action by the relation \cite{davies}
\begin{equation}
\langle T_{\nu }^{\mu }\left[ g_{\alpha \beta }\right] \rangle _{{\rm ren}%
}^{(0)}=2g^{\mu \sigma }(x)\frac{\delta }{\delta g^{\nu \sigma }(x)}W_{{\rm %
div}}[g_{\alpha \beta }].  \label{gravemt}
\end{equation}
Note that in odd spacetime dimensions the conformal anomaly is
absent and the corresponding gravitational part vanishes:
\begin{equation}
\langle T_{\nu }^{\mu }\left[ g_{\alpha \beta }\right] \rangle _{{\rm ren}%
}^{(0)}=0,\quad {\rm for\;even}\;D.  \label{gravemteven}
\end{equation}
The boundary part in Eq. (\ref{emtcurved1}) is related to the
corresponding flat spacetime counterpart (\ref{emtdiag}) by the
relation \cite{davies}
\begin{equation}
\langle T_{\nu }^{\mu }\left[ g_{\alpha \beta }\right] \rangle _{{\rm ren}%
}^{(b)}=\frac{1}{\sqrt{|g|}}\langle \bar{T}_{\nu }^{\mu }\left[
\eta _{\alpha \beta }\right] \rangle _{{\rm ren}}.
\label{translaw}
\end{equation}
By taking into account Eq. (\ref{emtdiag}) from here we obtain
\begin{equation}
\langle T_{\nu }^{\mu }\left[ g_{\alpha \beta }\right] \rangle _{{\rm ren}%
}^{(b)}=e^{(D+1)\sigma (r)}%
{\rm diag}(\varepsilon ,-p_{1},-p_{2},-p_{3},\ldots ,-p_{D}) ,
\label{bpartemt}
\end{equation}
Now we see that as gravitational part (\ref{emtcurved1}) is a
continous function on $r$ it does not contribute to the forces
acting on the
boundary and the vacuum force per unit surface acting on the boundary at $%
r=a$ is determined by the boundary part of the vacuum pressure, $%
p_{D}=-\langle T_{D}^{D}\left[ g_{\alpha \beta }\right] \rangle _{{\rm ren}%
}^{(b)}$, taken at the point $r=a$:
\begin{equation}
p_{D}(a)=e^{(D+1)\sigma (a)}F, \label{vacforce}
\end{equation}
where $F$ is given by (\ref{Fgum}).
\section{Casimir stress on cylindrical shell in dS background}
We will consider one of the simplest field-theoretical model in
which the domain wall type solutions appear \cite{viel}. The model
involves a single, real-valued scalar field $\varphi$ with
lagrangian given by
\begin{equation}
L=-1/2
g_{\mu\nu}\partial^{\mu}\varphi\partial^{\nu}\varphi-V(\varphi),\label{leq}
\end{equation}
and
\begin{equation}
V(\varphi)=\frac{\lambda}{2}( \varphi^{2}-v^{2})^{2},\label{veq}
\end{equation}
where $\lambda$ and $v$ are positive constants. The classical
ground states are given by $\varphi=\pm v$. The domain wall arises
if there are regions in the space where the field $\varphi$ has
different vacuum values, the domain wall interpolating between
such regions. In this paper we will consider a domain wall between
a cylindrical region around $z$ axis in which
$\varphi=\Lambda_{in}$ and the remaining part of the space where
$\varphi=\Lambda_{out}$.\\
  As an application of the general formulae
from the section-2 here we consider the important special case of
the dS$_{3+1}$ bulk for which
\begin{equation}
ds^{2}=\frac{\alpha^{2}}{\eta^{2}}[d\eta^{2}-\sum_{i=1}^{3}(dx^{i})
^{2}],\label{met}
\end{equation}
where $\eta$, is the conformal time
\begin{equation}
-\infty <\eta < 0.
\end{equation}
The constant $\alpha$ is related to the cosmological constant as
\begin{equation}
\alpha^{2}=\frac{3}{\Lambda}.
\end{equation}
Now we consider the pure effect of vacuum polarization due to the
gravitational field without any boundary conditions (to see such
problem for spherical shell and parallel plate geometry refer to
\cite{{set2},{set3},{set4}}). The renormalized stress tensor for
massless scalar field in de Sitter space is given by
\cite{{davies},{dow}}
\begin{equation}
< T^{\nu}_{\mu}>=\frac{1}{960
\pi^{2}\alpha^{4}}\delta^{\nu}_{\mu}.\label{grav}
\end{equation}
The corresponding effective pressure is
\begin{equation}
P=-< T^{1}_{1}>=-< T^{r}_{r}>=-\frac{1}{960
\pi^{2}\alpha^{4}},\label{peq}
\end{equation}
valid for both inside and outside the cylinder. Hence the
effective force on the cylinder due to the gravitational vacuum polarization is zero. \\
Now, assume there are different vacuum inside and outside
corresponding to $\alpha_{in}$ and $\alpha_{out}$ for the metric
Eq.(\ref{met}). Now, the effective pressure created by
gravitational part Eq.(\ref{peq}), is different for different part
of space-time
\begin{equation}
P_{in}=-< T^{r}_{r}>_{in}=-\frac{1}{960
\pi^{2}\alpha^{4}_{in}}=\frac{-\Lambda_{in}^{2}}{8640 \pi^{2}},
\end{equation}
\begin{equation}
P_{out}=-< T^{r}_{r}>_{out}=-\frac{1}{960
\pi^{2}\alpha^{4}_{out}}=\frac{-\Lambda_{out}^{2}}{8640 \pi^{2}}.
\end{equation}
Therefore the gravitational pressure over shell, $P_{g}$, is given
by
\begin{equation}
P_{g}=P_{in}-P_{out}=\frac{-1}{8640 \pi^{2}}(
\Lambda_{in}^{2}-\Lambda_{out}^{2})
\end{equation}
Now we considering the effective pressure due to the boundary
condition. Under the conformal transformation in four dimensions
with the conformal factor given by
  \begin{equation}
  \Omega(\eta)=\frac{\alpha}{\eta}.
  \end{equation}
The vacuum force acting from inside per unit surface of the
cylinder can be found using the Eqs. (\ref{qinsub}),
(\ref{vacforce}) for the vacuum radial pressure:
\begin{equation}
F_{{\rm in}}=\frac{\eta^{4}}{\alpha^{4}_{in}}p_{1}\mid
_{r=a-0}=\frac{\eta^{4}}{\alpha^{4}_{in}}
\frac{1}{4\pi^{2}a^{4}}%
\sum_{n=-\infty }^{+\infty }\int_{0}^{\infty }dz\, z
^{6}\frac{\bar{K}_{n}(z)}{\bar{I}_{n}(z)}%
F_{n}^{(p_{1})}[I_{n}(z)],  \label{Fin}
\end{equation}
with notation (\ref{Fnp1in}).
 The expression for the radial projection of the vacuum
force acting per unit surface of the cylinder from the outside
directly follows from Eqs. (\ref{qsubout}),(\ref{vacforce})
 with $q=p_{1}$:
\begin{equation}
F_{{\rm ext}}=-\frac{\eta^{4}}{\alpha^{4}_{out}}p_{1}\mid
_{r=a+0}=-\frac{\eta^{4}}{\alpha^{4}_{out}}\frac{1}{4\pi^{2}a^{4}}%
\sum_{n=-\infty }^{+\infty }\int_{0}^{\infty }dz\, z ^{6}
\frac{\bar{I}_{n}(z)}{\bar{K}%
_{n}(z)}F_{n}^{(p_{1})}[K_{n}(z)],  \label{Fext}
\end{equation}

Therefore the vacuum pressure due to the boundary condition acting
on the cylinder is given by \bea
 P_{b}&=&F_{{\rm in}}+F_{{\rm ext}}=\frac{\eta^{4}}{\alpha^{4}_{in}}
\frac{1}{4\pi^{2}a^{4}}%
\sum_{n=-\infty }^{+\infty }\int_{0}^{\infty }dz\, z
^{6}\frac{\bar{K}_{n}(z)}{\bar{I}_{n}(z)}%
F_{n}^{(p_{1})}[I_{n}(z)]
 \\
&-&\frac{\eta^{4}}{\alpha^{4}_{out}}\frac{1}{4\pi^{2}a^{4}}%
\sum_{n=-\infty }^{+\infty }\int_{0}^{\infty }dz\, z ^{6}
\frac{\bar{I}_{n}(z)}{\bar{K}%
_{n}(z)}F_{n}^{(p_{1})}[K_{n}(z)]\label{bunf} .\nonumber
 \eea
  The total pressure on the
cylinder, $P$, is then given by \bea
 P&=& P_{g}+P_{b}=\frac{1}{8640
\pi^{2}}( \Lambda_{out}^{2}-\Lambda_{in}^{2})+\frac{\eta^{4}}{36\pi^{2}a^{4}}%
\sum_{n=-\infty }^{+\infty }\int_{0}^{\infty }dz\, z
^{6}\label{ptot} \\
&&(\Lambda_{in}^{2}\frac{\bar{K}_{n}(z)}{\bar{I}_{n}(z)}%
F_{n}^{(p_{1})}[I_{n}(z)]-\Lambda_{out}^{2}\frac{\bar{I}_{n}(z)}{\bar{K}%
_{n}(z)}F_{n}^{(p_{1})}[K_{n}(z)])
.\nonumber \eea The
$\eta$- or time-dependence of the pressure in intuitively clear
due to the time dependence of the physical radius of cylinder.
This pressure corresponds to the attractive/repulsive force on the shell if $%
P</>0$. The equilibrium state for the cylinder correspond to the
zero values of Eq. (\ref{ptot}): $p=0$.
 Total pressure, may be negative or positive, depending on the
difference between the cosmological constant in the two parts of
space-time. Given a false vacuum inside of the cylinder  , and
true vacuum outside, i.e. $\Lambda_{in}> \Lambda_{out}$, then the
gravitational part is negative, and tends to contract the
cylinder, but the boundary pressure part may be positive or
negative. Therefore the total effective pressure on the cylinder
may be negative, leading to a contraction of the cylinder. The
contraction however, ends for a minimum of radius of the cylinder,
where both part of the total pressure are equal. For the case of
true vacuum inside the cylinder and false vacuum outside, i.e
$\Lambda_{in}< \Lambda_{out}$, the gravitational pressure is
positive. In this case, boundary part can be negative or positive
depending on the difference between $F_{in}$ and $F_{out}$ .
Hence, the total pressure may be either negative or positive.
\section{Conclusion}
In the present paper we have investigated the Casimir effect due
to the conformally coupled massless scalar field for a cylindrical
shell on background of the conformally-flat spacetimes. The
general case of the mixed boundary conditions is considered. The
vacuum expectation values of the energy-momentum tensor are
derived from the corresponding flat spacetime results by using the
conformal properties of the problem. Then we consider cylindrical
shell with constant comoving radius having different vacuums
inside and outside, i.e. with false/true vacuum  inside/outside.
The boundary induced part for the vacuum energy-momentum tensor is
given by Eq.(\ref{bpartemt}), and the corresponding vacuum forces
acting per unit surface of the shell have the form Eqs.
(\ref{Fin}),(\ref{Fext}).The effective vacuum pressure due to the
boundary condition acting on the cylinder is given by Eq.(37). The
vacuum polarization due to the gravitational field, without any
boundary conditions is given by Eq.(\ref{grav}), the corresponding
gravitational pressure part has the form Eq.(\ref{peq}), which is
the same from both sides of the shell, and hence leads to zero
effective force. However when we consider different cosmological
constants for the space between and outside of the shell, in this
case the effective pressure created by gravitational part is
 different for different part of the space-time and add to the
 boundary part pressure. The total pressure is given by
 Eq.(\ref{ptot}). Our calculation
 shows that for the cylindrical shell with false vacuum inside and
 true vacuum outside $\Lambda_{in}>
 \Lambda_{out}$, the gravitational pressure part is negative,
 but the boundary pressure part may be positive or negative. In contrast for the
 case of true vacuum inside the cylinder and false vacuum outside, $\Lambda_{out}>
 \Lambda_{in}$, the gravitational pressure is positive, and boundary part can be
 negative or positive depending on the difference between
 $F_{in}$ and $F_{out}$ in Eq.(\ref{bunf}). Therefore the detail dynamics of the
  cylindrical shell depends on different parameters and all cases of contraction and expansion
  may appear. The result may be of interest in the case of
 formation of the cosmic cylindrical domain walls in early
 universe.


\vspace{3mm}









\begin{thebibliography}{99}



 \bibitem {mueller}
 G. Plunien, B. Mueller, W. Greiner, Phys. Rep. {\bf 134}, 87(1986).
\bibitem{Moste}V. M. Mostepanenko and N. N. Trunov. The Casimir effect
 and its applications. (Oxford Science Publications New York,
 1997).
 \bibitem{Casimir}H. B. G. Casimir, proc. K. Ned. Akad. Wet. 51,
 793(1948).
 \bibitem{Remeo}E. Elizalde, S. D. Odintsov, A. Romeo, A. A. Bytsenko and
 S. Zerbini, zeta regularization techniques with applications(World
 Scientific, Singapore, 1994).
 \bibitem{Elizalde}E. Elizalde, Ten physical applications of
 spectral zeta functions, lecture notes in
 physics, (Springer-Verlage, Berlin, 1995).
\bibitem{mil3}K.A. Milton, {\it %
The Casimir Effect: Physical Manifestations of the Zero-Point
Energy}, hep-th/9901011; {\it Dimensional and Dynamical Aspects of
the Casimir Effect: Understanding the Reality and Significance of
Vaccum Energy}, hep-th/0009173.
 \bibitem{bord1}M. Bordag, U. Mohideen, V.M. Mostepanenko, Phys.
 Rept. {\bf 353}, 1-250, (2001).
 \bibitem{Mil} L. L. De Raad Jr and K. A. Milton, Ann. Phys.
 {\bf 136}, 2290, (1981).
 \bibitem{Gos} P. Gosdzinsky and A. Romeo, Phys. Lett. {\bf B441}, 265,(1998).
 \bibitem{Mil1}K. A. Milton, A. V. Nesterenko and V. V.
 Nesterenko, Phys. Rev. {\bf D59}, 105009, (1999).
\bibitem{zel}Ya. B. Zel'dovich, I. Yu. Kobzarev and L. B. Okun,
 Sov. Phys. JETP 40, 1 (1975).
 \bibitem{kib}T. W. B. Kibble, J. Phys. {\bf A9}, 1387 (1976).
 \bibitem{viel}A. Vilenkin, Phys. Reports {\bf 121}, 263 (1985).
\bibitem{hu} R. P. Huebener, Magnetic Flux Structure in
Superconductor (Spriner-Verlag, Berlin, 1979).
\bibitem{do} R. J. Donnelly, Quantized Vortics in Helium II
(Cambridge University Press, Cambridge, England, 1991).
\bibitem{ch} S. Chandrasekhar and G. S. Ranganath, Adv. Phys.
{\bf 35}, 507 (1986).
\bibitem{ar1} H. Arodz and A. L. Larsen. Phys. Rev. {\bf D49},
4154, (1994).
\bibitem{ar2} H. Arodz, Phys. Rev. {\bf D52}, 1082, (1995).
\bibitem{ba}M. Baker, J. S. Ball, and F. Zachariasen ,Phys. Rep.
{\bf 209}, 73 (1991).

\bibitem{hil} C. T. Hill, D. N. Schramm, and J. N. Fry, Comments.
Nucl. Part. Phys. {\bf 19}, 25 (1998).
 \bibitem{set} M. R. Setare and A. A. Saharian. Int. J. Mod.
 Phys. {\bf A16}, 1463(2001).
 \bibitem{set1}M. R. Setare and A. H. Rezaeian. Mod. Phys. Lett.
 {\bf A15}, 2159(2000).
 \bibitem{Ant}F. Antonsen and K. Borman. Casimir Driven Evolution of
 the Universe. gr-qc/9802013
\bibitem{set2}M. R. Setare and R. Mansouri. Class.Quant.Grav. 18 (2001) 2331
 \bibitem{set3}M. R. Setare. Class.Quant.Grav. 18 (2001) 4823-4830

\bibitem{set4}M. R. Setare and R. Mansouri.Class.Quant.Grav. 18 (2001) 2659-2664


 \bibitem{vil1}A. Vilenkin and E. P. S. Shellard. COSMIC STRINGS
 AND OTHER TOPOLOGICAL DEFFECTS,(Cambridge University Press,
 1994).
\bibitem{RomSah}  A. Romeo and A. A. Saharian, J. Phys. {\bf A35}, 1297,
(2002).
\bibitem{Saha01a}  A. A. Saharian, Phys. Rev. {\bf D63}, 125007 (2001).

\bibitem{Saha01b}  A. Romeo and A. A. Saharian, Phys. Rev. {\bf D63}, 105019
(2001).
\bibitem{davies}N. D. Birrell and P. C. W. Davies, Quantum fields in curved
  space,( Chambridge University press, 1986).

 \bibitem{dow} J. S. Dowker, R. Critchley, Phys. Rev. {\bf D13},
 3224 (1976).


\end{thebibliography}
\end{document}